\documentclass{article}
  \usepackage{graphicx}
\begin{document}

\title{Special collision dynamics of massless fermions leading to confinement}
\author{J.Witters\\Department of Physics,University of
Leuven, Belgium}
\maketitle
\begin{abstract}
This is a theory for the fundamental structure of nature based on
spinors as unique building blocks.
Confinement of the spinors in particles is due uniquely to the
dynamics of the collisions between them. The interaction operator
contains only spin and momentum operators. The eigenfunctions of the interaction operator have negative as well as positive eigenvalues, limited in magnitude to the Planck energy. Interactions in one-center particles between spinors with eigenvalues of different sign tend to bring the energy of the particle closer to zero.  At the origin of the
energy reduction through interaction is the spin precession,
taking place when two spinors of different type come together.
The model can explain the existence of stable particles and all
the known forces, including gravitation.
\end {abstract}
\section* {introduction}
The rules of wave mechanics demand large wavenumbers for
particles which are confined in small volumes. Particles like the
electron or the quarks, which are smaller than any measurable
size, should therefore have a much larger rest energy than they
actually have. In the existing theories up to now, fields are
introduced to provide negative potential energy. To make that
work, an artificial distinction has to be made between real
particles and virtual field-carrying particles. The particles of
the fields are allowed to disobey the rules which they are
supposed to save. This is an unsatisfactory situation.
\\ We will show that the rules can be saved in a different way,
without the artificial introduction of fields. To meet this goal
we introduce an interaction operator based only on spin-and
momentum operators, and let it operate on two-vectors. Accepting
the eigenstates of this operator which have negative as well as
positive eigenvalues, we arrive at two different types of spinors
with collision dynamics which lead to confinement. Accepting the spinors with negative eigenvalues as real does not lead to a radiation catastrophe because radiation brings the energy of the emitting particle to zero. The energy ladder of a one-center particle does not extend to minus infinity but is limited in magnitude by the Planck energy. All the known
particles  are formed by local interactions between
the spinors which constitute them. \\
This gives us surprising results. For example the
result that gravitation is just another manifestation of
electromagnetism.

\section{The fundamental Hamiltonian and some commutation relations.}
We will base our construction of physical reality on the
following interaction operator $H$, which we will call an
Hamiltonian, although it can not in general be diagonalised when
acting on localized particles :\begin {equation}H
=c(\hat{\sigma}.\vec{p})\end {equation} where c is the velocity
of light, $\hat{\sigma}$ is a vector representation of Pauli spin
matrices and $\vec{p}$ is the momentum operator
$-i\hbar\vec{\nabla}$ = $ -i\hbar\vec{grad} $.
\\ The states on which $H$ operates are two-vectors $\psi$:
\begin {equation} \psi =\left(\begin{array}{c}
    f(\vec{r},t)  \\
  g(\vec{r},t)
\end{array}\right) \end {equation}
 We will leave the time dependence out of the functions further on because that can be
deduced from motions on the light cone, putting $e^{i\omega t}$
equal to $e^{ickt}$ .\\ Plane wave spinor eigenfunctions of $H$
are $\psi_{+}$ and $\psi_{-}$:
\begin {equation}\psi_{+}=\left( \begin{array}{c}
    N\exp{i\vec{k}.\vec{r}}  \\
 0
\end{array}\right)\\ \; \; \; \; \;   \;  \psi_{-}=\left(\begin{array}{c}
 0\\   N\exp{i\vec{k}.\vec{r}}
\end{array}\right)
 \end {equation}
Where N is a normalisation constant and spins are quantised in
the direction of $\vec{k}$ . The eigenvalues are $+\hbar k$ and
$-\hbar k$ respectively.\\ \setlength{\unitlength}{1mm}
\begin{picture}(200,12)(0,0)
\put(50,7){\vector(-1,0){40}}\put(55,7){\vector(1,0){40}}\put(40,6){\vector(-1,0){10}}\put(70,6){\vector(-1,0){10}}\put(30,2){spin}\put(65,2){spin}
\put(15,2){$\psi_{+}$}\put(80,2){$\psi_{-}$}\put(5,7){$\vec{k}$}\put(100,7){$\vec{k}$}
\end{picture}
$H$ acts exclusively on sums of spinors which move with the speed
of light. This can be concluded by determining the velocity
operator $\frac{d\vec{r}}{dt}$:
\begin {equation} \frac{d\vec{r}}{dt}= \frac{i}{\hbar}
[\vec{r},H] = c\hat{\sigma} \end{equation}The motion of fermions
with the velocity of light was recognized long ago, soon after
the introduction of the theory of Dirac for relativistic
fermions. It was called "Zitterbewegung" [1] ,was interpreted as an
irregular motion due to uncertainty, and was then neglected. It
is a central point in this paper.\\ It is important to note that
the motion of a spinor is along the spin vector, not along the
momentum vector. The motion of the center of mass, however, is
along the momentum vector, as it should be. This is deduced from
the definition of the position vector $\vec{r}_{M}$ of the center
of mass and from the commutation of $\vec{r}_{M}$ with $H$:
\begin {equation} \vec{r}_{M} = \frac{\vec{r}H}{<H>} \end {equation}
This definition is the operator form of the classical formula
$\frac {\sum \vec{r}_{i}m_{i}}{\sum  m_{i} }$.\\ $<H>$ is the
expectation value of $H$, integrated over a limited space
containing all the spinors which make up the one-center particle under
consideration.
\\ The motion of the center of mass is determined by:
\begin {equation}\frac{d \vec{r}_{M}}{d t} =\frac{i}{\hbar} \frac{[\vec{r}H,H]}{<H>}=\frac{c^{2}\vec{p}}{<H>}=\frac{c^{2}\vec{p}}{m c^{2}}=\frac{\vec{p}}{m} \end {equation}
We have left out a vector product of spin and momentum in this result. This is justified for non-interacting spinors, with spins quantised in the direction of $\vec{k}$. In the full expression for interacting spinors in particles the sum of the vector product terms is zero on account of the central symmetry. The result in equation (6) is as we expect from classical experience.\\ Commutation
relations for $\hat{\sigma}$, for $\vec{p}$ and for
[($\vec{r}\times \vec{p}) + \frac{\hbar}{2}\hat{\sigma} $] give
us the following results:
\begin {equation}\frac{d \hat{\sigma}}{d t} =\frac{i}{\hbar}
[\hat{\sigma} ,H]=\frac{ 2 c}{\hbar} \hat{\sigma} \times \vec{p}
\; \; \;\;\;\;\;\; \frac{d \vec{p}}{d t} = 0\;\;
\;\;\;\;\;\;\frac{d [ (\vec{r} \times \vec{p}) +
\frac{\hbar}{2}\hat{\sigma}]}{d t} = 0
\end{equation} The first equation in (7) is the key equation to understand the
origin of confinement without fields, as will be shown in the
next section.\\ If  $<H>$ is calculated for one isolated
particle, in which all the spinors follow closed paths around one
center, the result may be negative as well as positive, depending
on which sign of the spin-momentum product prevails. However, we may keep the convention that $|<H>|$, the norm of
$<H>$, is the energy of the particle.
\\The sign of $<H>$ can be identified with the sign of the electric charge.
\\ The difference between the theory which we present here and
the Dirac theory [1,2] for relativistic fermions lies mainly in the
introduction of lower-level structures and in the acceptance of
negative values for $<H>$. These were rejected as unphysical because of a "radiation catastrophe". In our model all energy levels are brought to zero by interactions, not to minus infinity. \\ The Dirac Hamiltonian [1,2] is an approximation of $H$, in
the following way:
\begin{equation} H = c\hat{\sigma}.\vec{p}
=c\hat{\sigma}.\vec{p}_{M}+ c\hat{\sigma}.(\vec{p}-\vec{p}_{M})
\end{equation}
$\vec{p}_{M}$ represents the collective momentum of the whole
particle, deduced from the motion of the center of mass. The
Dirac Hamiltonian $H_{D}$ = $c \hat{\alpha}.\vec{p} + m_{o}
c^{2}\hat{\beta}$ with $\hat{\alpha}$ and $\hat{\beta}$
four-by-four matrices built out of $\hat{\sigma}$ and unit
matrices is a kind of renormalised form of equation (1), with
$\vec{p}$ representing the center-of-mass momentum and the rest
energy $ m_{o} c^{2}$ corresponding to $|<H>|$ in the
center-of-mass frame. By taking this definition of $\vec{p}$ ,
the energy-reducing interactions were lost in $H_{D}$. Also the
$\hat{\sigma}$ submatrices of $\hat{\alpha}$ in the Dirac theory
do not refer to the spin operator acting on spinors but rather to
an averaged angular momentum operator for the whole particle.
\section{Collision dynamics}
Spinors interact with one another only locally, because the
interaction operator $H$ is a local operator. As shown in
equation (7), $\hat{\sigma}$ rotates in precession around
$\vec{p}$, much like a magnetic moment precesses around a
magnetic field. As long as the spinors are separated in
space-time, the spins are aligned with the momenta. When spinors
come closer than a minimum distance $r_{o}$, defined further on,
the spins and the momenta are added up and the total spin will
rotate around the total momentum.\\ The total momentum and also
$\hat{\sigma}^{2}$ are conserved in the collisions. Collisions
between spinors with charges of the same sign are fairly
classical. The
outgoing vectors of spin and momentum are simply rotated with
respect to the incoming vectors, rotated around the total
momentum.
\\
Collisions of spinors with different charge signs, however, are
very special. They have no similarity to what we know in the
macroscopic world. They are at the origin of confinement in
localized particles. The spin precession produces here a kind of
bending of the outgoing tracks towards one another, or, if we
keep all the vectors in one plane, a reflection of the motion on
the line of the total momentum. To see that we must remember that
the direction of motion is along the spins. Two examples are
shown in figure (1):\\ \begin{center}
   \includegraphics*[width=0.95 \textwidth]{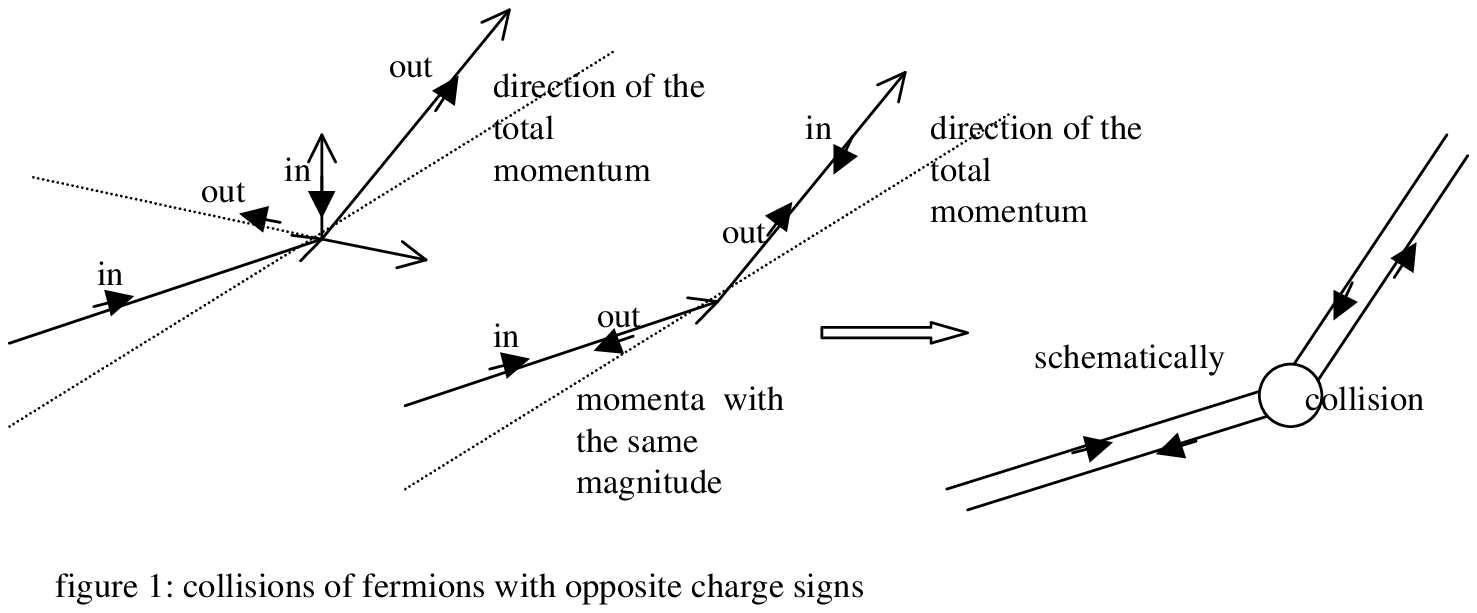}
\end{center}
 On tracks bent by this kind of collisions, four spinors can run
 around the sides of a square and many intertwined spinors can form
  circular paths. \\ There is no need to introduce a field to bend the
  tracks. The field is the other spinor. Because spinors  with
  different charges stay together and are transformed into
each other, the energy of a particle is reduced as a consequence
of periodical switching between positive and negative levels.
\section{The photon}
A neutral combination of fermionic spinors with cylindrical
symmetry has all the properties of the particles which we know as
photons. In a photon, the total angular momentum with respect to
the axis of the cylinder, which is the axis of propagation, is
constant. It is composed of a spin part and an orbital part, both
having half integer quantum numbers. Cylindrical spinor functions
for a particle which is stable under the interaction operator $H$
must be eigenfunctions of $H^{2}$, i.e. of the $\vec{\nabla}^{2}$
operator. \\ We can read in handbooks on mathematical functions
that the eigenfunctions of the $\vec{\nabla}^{2}$ operator which
are finite on the axis are the cylindrical Bessel functions. In
cylindrical coordinates with the axis of the cylinder along z, the radius r and the azimuthal angle $\varphi$ are defined in the xy plane and the lowest order cylindrical Bessel functions are:
\begin{equation}
u_{+}(z,r,\varphi)=Ne^{ikz}\frac{sin(k_{o}r)}{\sqrt{k_{o}r}}e^{i\frac{\varphi}{2}}
\; \; \; \; \; \;
v_{+}(z,r,\varphi)=Ne^{ikz}\left(\frac{\cos(k_{o}r)}{\sqrt{k_{o}r}}-\frac{\sin(k_{o}r)}{\sqrt{(k_{o}r)^{3}}}\right)e^{3i\frac{\varphi}{2}}
\end{equation}
and
\begin{equation}
u_{-}(z,r,\varphi)=Ne^{ikz}\left(\frac{\cos(k_{o}r)}{\sqrt{k_{o}r}}-\frac{sin(k_{o}r)}{\sqrt{(k_{o}r)^{3}}}\right)e^{-3i\frac{\varphi}{2}}
\; \; \; \; \; \;
v_{-}(z,r,\varphi)=Ne^{ikz}\frac{sin(k_{o}r)}{\sqrt{k_{o}r}}e^{-i\frac{\varphi}{2}}
\end{equation}
The parameter $k_{o}$ in these functions must be chosen to match
the smallest meaningful distance $r_{o}$ so that $k_{o}r_{o}=1$.
We will take the Planck length [3] $\simeq 10^{-35}m$ for $r_{o}$.
\\ We can see that the following $\psi_{ph}$, with spin quantisation in the z-direction, is an eigenfunction
of $H^{2}$ with all the properties of the phonon:
\begin {equation}\psi_{\pm} =\left( \begin{array}{c}
   u_{\pm}   \\
   v_{\pm}
\end{array}\right) \end{equation}
 This spinor has standing waves in the radial direction which provide the necessary communication between the circular currents around the axis of the cylinder. It can transport at the same
 time a positive contribution to the energy of a detector and a
 negative contribution to the energy of a source, transmitting also an angular momentum $\hbar$. Detector and
 source are on the same footing; no "strange actions at a
 distance". With k equal to zero we have a vacuum photon, which
 carries no energy but can couple particles together.\\  We can
 depict the photon as a sort of hollow tube, with zero density on
 the axis, as shown in figure (2)
 \begin{center}
    \includegraphics*[width=0.95 \textwidth]{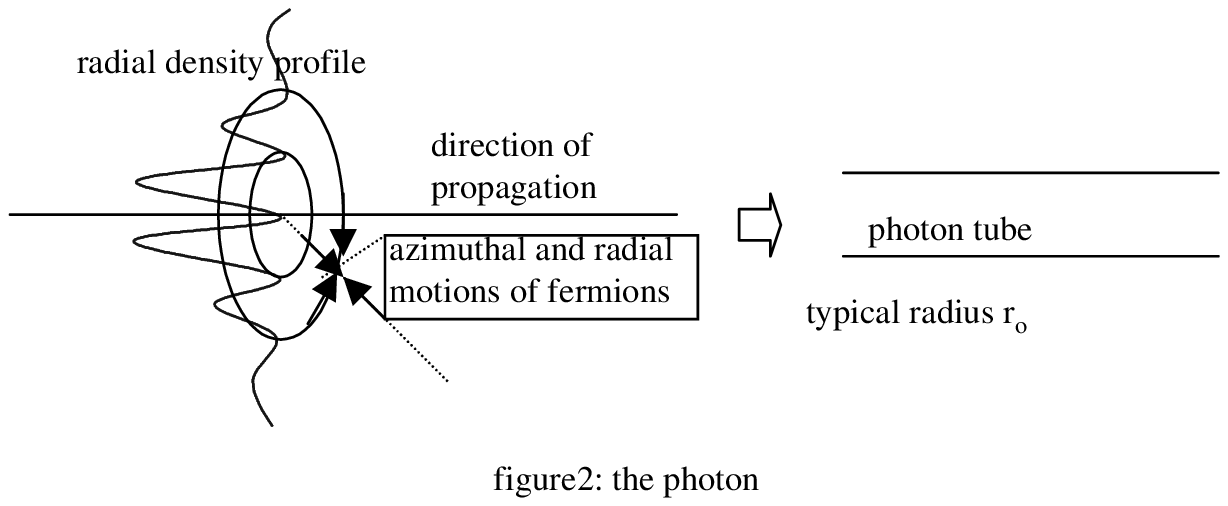}
\end{center}
When the photon is transmitting "force" between two charged
particles in a stationary bound state, it can be imagined as a
whirl with the axis in the most favorable direction for the
binding, that is perpendicular to the line connecting the
particles.
\section{Localised particles; the electron}
From here on the systems become too complicated for complete
analytical description. We will only be able to give rough
pictures, which will have to be refined with the aid of large
computers.\\ A spinor for one type of charge can not be radially
confined without reflections caused by the interaction with
photons. The most efficient mode to achieve radial confinement is
probably the internal reflection of a charged spinor moving
axially in the inside of a photon,as shown in figure (3)
   \begin{center}
      \includegraphics*[width=0.95 \textwidth]{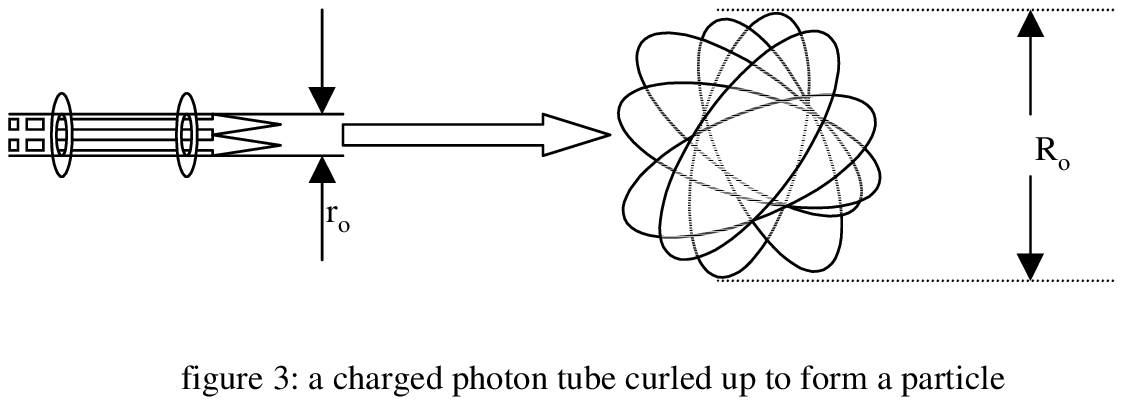}
  \end{center}
   We will call this a charged photon tube. We can imagine
internal reflections to prevail over outward scattering by
correlated radial motions of the charged interior of the tube and
of the components of the photon. The photon shield will probably
pick up energy in that process but reduce the energy of the whole
system. \\ Due to the interaction between the charged spinor and
the photon the cylindrical symmetry may be broken spontaneously.
The charged photon tube curls up and the cylinder symmetry
survives only locally. The tube may now close on itself and if
the conditions are right for self-consistency we may have a
confined fermion. It resembles a ball of yarn as shown in figure
(3). \\\\  The ball can be hollow to prevent the formation of too
high momentum components in the center. \\ The total length of
the woven charged tube must fit the condition that the path
integral of the phase gradient is equal to an integer times
2$\pi$ . Although the exact determination of the structure of
this ball will be a formidable task, we may try to guess at the
relation between the radius R of the ball and its energy. With
$k_{o}r_{o}=1$ and $r_{o}$ the Planck length [3] of $10^{-35}m$,
$\hbar k_{o}$ is of the order of 10Nm. The energy of one
elementary fermion circulating at a distance $r_{o}$ from a
center would then be of the order of $10^{9}J$. The energy of a
curled-up ball will be inversely proportional to the number of
Planck-size cells in the volume occupied by the charged current,
each cell being visited only once in a cycle to avoid
self-intersection. If the ball has a radius R and no hole, the
number of Planck-size cells is of the order of
$(\frac{R}{r_{o}})^{3}$, reducing the momentum of the fermion by
that factor. \\ The rest energy of the electron is $8.10^{-14}J$.
Putting this equal to $\frac{10^{9}.(r_{o})^{3}}{R^{3}}$ we find
R equal to approximately $10 ^{-28}m$ or $10 ^{7}$ Planck
lengths. In this estimation we have completely neglected the
contribution of the photon shield to the energy of the particle.
The more compensation given by the photon shield, the higher the
bare momentum of the charged spinor and the lower the radius of
the particle. On the other hand the radius of the electron may be
bigger if there is a central hole. This would not necessarily
show up in collisions with other particles because the hole would
make the ball softer.
\\ Particles like the quarks, with an energy which is rather
close to the energy of the electron on the Planck scale, both
being relatively very small, can be similar to the electron.
\section{The interactions}
The interactions between particles is mediated by photons. In a
simplified scheme we take two different particles with momenta
$p_{1p}$ and $p_{2p}$ respectively, with the centers at a
distance $r$ from one another. A photon whirl makes a connection
between the two and can take up momentum components
simultaneously from both currents, as shown in figure (4).
  \begin{center}
     \includegraphics*[width=0.95 \textwidth]{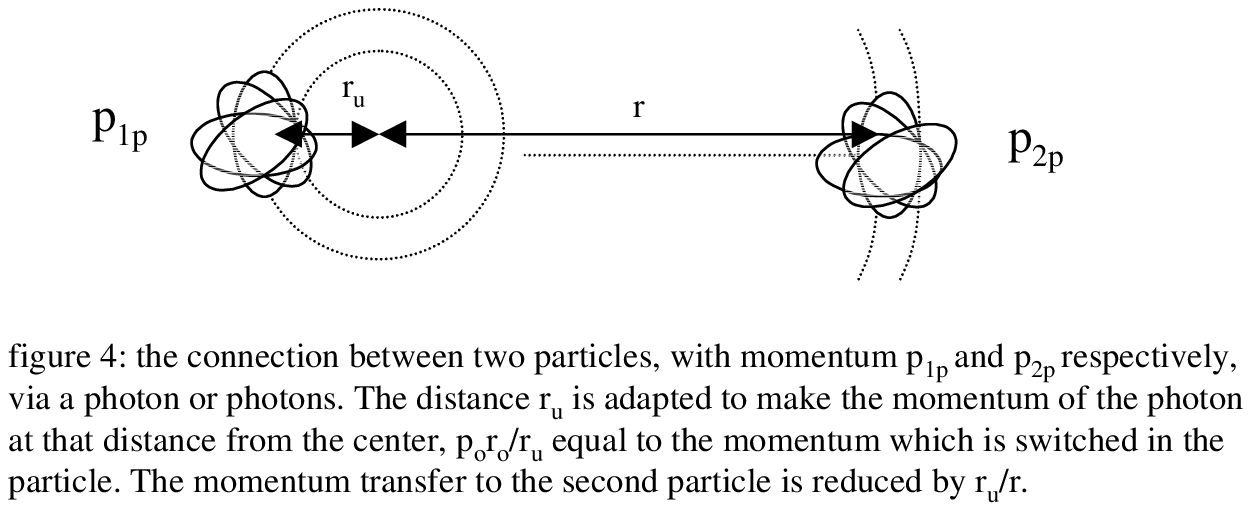}
 \end{center}
 The system will radiate out energy if it is
not in a stable state. In a stable situation, in which every
spinor in the system returns periodically to a given initial
condition, the photon whirl receives simultaneously from both
particles equal and opposite spinor components which it takes up
without changing its energy. Each of the particles is subjected,
via collisions with the photon whirl, to a reduction of its
energy with respect to the energy it would have without the
interaction. This reduction can be estimated by considering a
match of the momentum vectors of the photon with the momentum of
the particle on one side, say $p_{1p}$. To make this match we
must consider the bare momentum  $p_{u}$ of the spinor unit with
which the photon interacts. For a charged particle like the
electron this is the bare momentum of the central spinor. For
interaction between bosons which carry energy it is the bare
spinor momentum $p_{o}=\hbar k_{o}$. If the photon is positioned
at a distance $r_{u}= \frac{\hbar}{p_{u}}$ from the first
particle, in a way to get maximum contact interaction, it can
contribute a fraction $\frac{r_{u}}{r}$ at the position of the
second particle to reduce the energy of the system. Putting this
together in a formula for the interactive reduction of energy we
arrive at: \begin{equation}interactive\; \; energy \; \;
reduction \; \; = c \frac{p_{1p}p_{2p}}{p_{u}}\frac{r_{u}}{r}=
c\hbar \frac{p_{1p}p_{2p}}{(p_{u}^{2}) r} \end{equation} with
$cp_{1p}$ and $cp_{2p}$ the energies of the interacting particles
when they are isolated,  $p_{u}=\frac{\hbar}{r_{u}}$ the momentum
of the bare spinor unit with which the photon couples and r the
distance between the particles.\\ If we apply this perturbation
energy correction to charged particles, $p_{1p}$ and $p_{2p}$ are
lower than ${p_{u}}$ by the amount of compensating momentum in
the photon shields of the particles.\\  We know from the value of
the fine structure constant $\frac{1}{137}$ that by putting
\begin{equation}electrical\; \; interaction\; \; \frac{e^{2}}{4\pi \epsilon_{o} r}=
c\hbar \frac{p_{1p}p_{2p}}{(p_{u}^{2}) r} \end{equation} the
reduction factor $\frac{p_{1p}p_{2p}}{(p_{u}^{2})}$ must be
$\frac{1}{137}$. \\ Gravitation is an interaction between bosons.
In this interaction the mediating photons couple two-by-two the
elementary spinors which are not completely cancelling each
others energy in the particles. Here we must take the bare
$p_{o}=\hbar k_{o}$ momentum as the unit for interaction with the
photons. Taking for the momenta which must be compensated $m_{1}
c$ and $m_{2} c$, deduced from the energies of the particles, we
arrive at:\begin{equation}gravitational \; \; interaction \; \;
\frac{ Gm_{1}m_{2}}{r} = c\hbar \frac{c m_{1}cm_{2}}{(p_{o})^{2}
r} =  c \frac{c m_{1}cm_{2}(r_{o})^{2}}{\hbar r}
\end{equation} and this leads to the value of the Planck length when the known value of the gravitational constant G is inserted.
\\ The interaction of particles can also give short-range
forces, like the strong and the weak nuclear force. The short
range of those forces points towards a rupture of the bond
between particles. The obvious candidate for these forces is a
connection via charged photon tubes, entangled into the connected
particles. A particle like an electron , with a total length for
the charged photon tube of the order of $10^{-14}m$, partially
unwound and hooked into similar particles, gives us the right
order of magnitude for the range of the nuclear forces. \\
 \\ 

\begin{thebibliography}{x}
\bibitem[1]{1} treated in most textbooks on relativistic quantum mechanics e.g. W.Greiner, Relativistic quantum mechanics; Springer Verlag 1990 ISBN3-540-50986-0
\bibitem[2]{2} P.A.M.Dirac, Proc.Roy.Soc.A117,610(1928);A118;351(1928) "The Principles of Quantum Mechanics"
\bibitem[3]{3} original reference: M.Planck, Sitz.Preuss.Acad.d.Wissensch., Berlin 1899, pp.440-470
\\
recent reference: "The role of Newton's constant in Einstein's gravity" by Vittorio de Alfaro; in "The High Energy Limit", A.Zichichi Ed., 1983 Plenum Press, Proc. 18th Course of the Int'l Course of Subnuclear Physics, 1980 Erice.
\end{thebibliography}
\end{document}